 \documentclass[prl,twocolumn]{revtex4}

\begin{document}
 
 \title{Torque Anomaly in Quantum Field Theory}
 
\author{S. A. Fulling}
\email{fulling@math.tamu.edu}
 \homepage{http://www.math.tamu.edu/~fulling}
 \affiliation{Departments of Mathematics and Physics, Texas A\&M 
University, 
  College Station, TX, 77843-3368 USA}

  \author{F. D. Mera}
\affiliation{Department of Mathematics, Texas A\&M 
University, 
  College Station, TX, 77843-3368 USA}

 \author{C. S. Trendafilova}
 \email{cyntrendafilova@gmail.com} 
\affiliation{Department of Physics, University of Texas,
 Austin, TX, 78712-0264    USA}
  
 \date{December 26, 2012}

\begin{abstract}
The expectation values of energy density and pressure of a quantum 
field inside a wedge-shaped region appear to violate the expected 
relationship between torque and total energy as a function of 
angle.  In particular, this is true of the well-known 
 Deutsch--Candelas stress tensor for the electromagnetic field, 
whose definition requires no regularization except possibly at the 
vertex.
Unlike a similar anomaly in the pressure exerted by a reflecting 
boundary against a perpendicular wall, this problem cannot be 
dismissed as an artifact of an ad hoc regularization.
  \end{abstract} 
 
 \maketitle

  \subsubsection{Introduction} \label{sec:intro}
The law of conservation of energy requires that the change in 
energy of a system, such as a fluid in a box,
  in response to an infinitesimal motion of an element of its 
  boundary be equal to the negative of the force on that boundary 
  times its perpendicular displacement.  Thus, for perpendicular 
  motion of a flat boundary at $x=\mathrm{const.}$ one has
 \begin{equation}
-\,{\partial{E}\over\partial x}  = F = \int\!\!\!\int p\, dy\,dz\,,
\label{pressurebal} \end{equation}
 where $E$ is the energy, $p$ is the pressure, and the integral is 
over the moving element of boundary.
 Similarly, for a wedge of opening angle $\alpha$ one expects the 
change in energy with respect to angle to be related to the torque 
on the moving side:
 \begin{equation}
  -\,{\partial E\over\partial\alpha} =  \tau 
 = \int\!\!\!\int rp\, dr\, dz
\label{torquebal} \end{equation}
in cylindrical coordinates.
 Such relations (sometimes called instances of  the ``principle of 
virtual work'') do not follow automatically from the 
 local energy-momentum conservation law, 
 ${\partial{T^{\mu\nu}}\over{\partial x^\mu}} =0\,$;  
 the equation of state of the 
matter (dependence of pressure on energy density) 
 must be consistent with the dependence of the energy density on 
the parameter concerned.

 Recent work on quantum vacuum energy has displayed violations of 
(\ref{pressurebal}) that have been traced to cutoffs introduced, 
without adequate physical basis, to remove divergences in the total 
energy near an idealized boundary.
 (See \cite{EFM} and references therein.)
 An ad hoc remedy was achieved (reviewed below).
 In our more recent work on wedges, it was expected that a similar 
problem and remedy would arise in connection with 
(\ref{torquebal}).
 As reported below, the problem arose but the remedy did not work.
 More importantly, we point out here that (\ref{torquebal}) is 
violated already for conformally invariant fields \cite{DC}, where 
there is no boundary divergence in a wedge (except at the axis), 
and hence the result cannot be blamed on a bad regularization 
method.

 \subsubsection{Classical fluid}\label{sec:classical}
 We begin by reviewing how (\ref{torquebal}) manifests itself for a 
classical fluid or a more general radially layered system.
Assume an equation of state $p=\beta\rho$, where, at first,  
 $\rho$ and $p$ are homogeneous 
(but depend on~$\alpha$).  
 Consider a wedge region ($0<\theta<\alpha$) with large outer 
radius~$R$, and assume that there is no shear stress $T^{r\theta}$
 on that outer boundary. All quantities are considered \emph{per 
unit length} in the $z$ direction 
 (e.g., the ``volume'' $V$ has units of area).  Then
$E = \rho V = {\textstyle\frac12}R^2\alpha\rho$, so
 \begin{equation}
   {\partial E\over \partial {\alpha} }
 ={\textstyle\frac12}R^2\left(\rho+\alpha\,{\partial {\rho}
 \over \partial {\alpha}}\right) .
 \label{startclas}\end{equation}
 Also, 
 \begin{equation}
 \tau= \int_0^R p \,r\,dr 
 =   {\textstyle\frac12}R^2\beta\rho.
 \end{equation}
 So we expect that  
 \begin{equation}\rho+\alpha\,
 {\partial \rho\over  \partial \alpha}
  = -\beta\rho,
 \label{fluidode}\end{equation}
  to satisfy (\ref{torquebal}).  
 Eq.~(\ref{fluidode}) implies  that
$ \rho \propto \alpha^{-(\beta+1)}$ and  hence $E$ has the form
 \begin{equation}E = c_1\alpha^{-\beta}
 =  c_2 V^{-\beta}.
\label{endclas} \end{equation}
The final formula is  shape-independent and \emph{equivalent}
  to the equation of state. 

 Since the torque balance holds locally for each~$r$,
  this discussion  generalizes to $\rho$,
 $p$, and $\beta$ dependent on $r$ (for example, to surface tension
 in a cylindrical membrane).

 \subsubsection{Rectilinear pressure anomaly and its resolution} 
\label{sec:pressanom}
 Next we review \cite{EFM}.  Consider a scalar field with the 
simplest curvature coupling, $\xi=\frac14$. (Other choices do not 
affect the situation significantly.)
 At distance $x$ from a perfectly reflecting plane boundary the 
expectation value of the energy density is 
$\rho = (32\pi^2 x^4)^{-1}$ 
and that of the pressure parallel to the  plane is the negative 
of that;
 therefore, on a test surface perpendicular to  the plane, the 
density 
and pressure satisfy (\ref{pressurebal}) pointwise, but the total 
energy and force (integrated over ~$x$) are divergent.
 Of course, a real boundary cannot be perfectly reflecting at 
arbitrarily high frequencies, and an arbitrary, but physically 
plausible, response is to insert an exponential ultraviolet cutoff.  
The resulting energy and force violate (\ref{pressurebal}) by a 
factor $-2$.  
[\emph{Note:} The rightmost member of Eq.~(20) in \cite{EFM} has 
the wrong sign.]
The ultraviolet cutoff is related to point-splitting 
in the time direction, so it is natural to consider splittings 
(by distance $\epsilon$) in 
the various space directions; choosing the remaining orthogonal 
space direction yields functions satisfying (\ref{pressurebal})
 (and, incidentally, agreeing with the ultraviolet regularization 
for the pressure, not for the energy):
 \begin{equation}
\rho= \frac1{2\pi^2(\epsilon^2+ 4x^2)^2} = - p.
 \end{equation}

 \subsubsection{Scalar quantum field theory in a 
wedge}\label{sec:cutofftheory}
 In \cite{T12} and \cite{Dowker} the cutoff
technologies of  \cite{EFM} and predecessor papers
 were applied in cones and wedges.  
 The obvious analog of the cutoff successful in the rectilinear 
case is point-splitting in the axial ($z$) direction.
 There were two surprises:   
 (1) The cutoff did not
completely remove the divergence at the axis  of the
wedge.  (2) The expected equality (\ref{torquebal}) 
 was not satisfied, even with the axial cutoff.
We intend to improve and extend our \emph{Mathematica} 
calculations to this effect before publishing any details.
Our attempts to resolve the conundrum about the torque were 
interrupted by 
the observation reported in the next section.

\subsubsection{The Deutsch--Candelas stress tensors for conformally 
invariant fields} \label{sec:conformal}
 In a classic paper \cite{DC}, whose results have been confirmed by 
independent calculations (e.g., \cite{BL}),
  Deutsch and Candelas calculated 
the energy density and pressure in a wedge for the conformally 
coupled ($\xi=\frac16$) scalar field and the electromagnetic field.
 For these fields there is no divergence against a flat boundary
 (though the divergence at the axis remains, and weaker divergences 
emerge for curved, smooth boundaries).
 Therefore, for strictly positive values of $r$ one can 
meaningfully study the unregularized quantities, and the issue of a 
trustworthy cutoff does not arise.

  According to \cite{DC},  the vacuum stress 
tensor inside a wedge, in coordinates $(t,r,\theta,z)$ and metric 
signature $g_{00}<0$, is \begin{equation} 
 T^\nu_\mu = 
{f(\alpha)\over 720 \pi^2r^4}\,
  \mathrm{diag}(1,1,-3,1)
\label{wedgestress}    \end{equation}
   with
 \begin{widetext}
 \begin{equation}
 f(\alpha)=
 \cases
{ \displaystyle {\pi^2\over 2\alpha^2}
 \left({\pi^2\over\alpha^2} - {\alpha^2\over \pi^2}\right)\\
&\hbox{for conformally coupled scalar field,} \cr
\displaystyle
  \left({\pi^2\over\alpha^2}+11\right) \left({\pi^2\over\alpha^2}-1\right)
&\hbox{for electromagnetic field.} 
 \cr}
 \end{equation}
 \end{widetext}
 Confining attention to a finite interval  
$r_\mathrm{min}<r< r_\mathrm{max}\,$, 
 consider  the torque on the side of the wedge at $\theta=\alpha$ (the 
other side remaining at $\theta=0$).
One may consider $r_\mathrm{max}/r_\mathrm{min}$ to be large, so 
that the region approximates a complete wedge, or to be small, to 
fix attention on a cylindrical layer of vacuum energy at one 
particular~$r$; the result is completely uniform in this 
parameter.
 From the pressure $T_\theta^\theta\,$, the torque is (per 
unit~$z$)
 \begin{equation} \tau = \int_{r_\mathrm{min} }^{r_\mathrm{max} }
     r \, dr 
 T_\theta^\theta  (r)
 =- \int_{r_\mathrm{min} }^{r_\mathrm{max} }    r \, dr 
 \,{1\over 720 \pi^2r^4} \, 3f( \alpha).
 \end{equation}
But we should also be able to calculate it from the 
 $\alpha$-derivative of the energy (per unit~$z$)
 \begin{eqnarray}
 E &= \displaystyle
  - \int_{r_\mathrm{min} }^{r_\mathrm{max} }r   \, dr 
  \int_{0}^{\alpha} \,  d\theta \, T_{00}\nonumber\\
 &= \displaystyle
 -\int_{r_\mathrm{min} }^{r_\mathrm{max} } r \, dr  \,
 {1\over 720 \pi^2r^4} \,  \alpha f(\alpha).
 \end{eqnarray}
 Thus
 \begin{equation}
  \tau = -\, \frac{\partial {E}   }{ \partial {\alpha}  }
 =   \int_{r_\mathrm{min} }^{r_\mathrm{max} } r \, dr  \,
 {1\over 720 \pi^2r^4} \, 
 \frac{d }{d \alpha } [  \alpha   f(\alpha )].
 \end{equation}
 Consistency will therefore be achieved if
 \begin{equation}g(\alpha) \equiv \frac{ d }{d \alpha}  [\alpha  f(\alpha)]
 +3f(\alpha) =0.
 \end{equation}
 We have 
\begin{equation}
 \alpha f(\alpha) = \cases{\textstyle
        \frac{1}{2}\, \pi^4 \alpha^{-3} -\frac{1}{2}\,\alpha
 &\hbox{scalar}, \cr \noalign{\smallskip}  \textstyle
   \pi^4\alpha^{-3} + 10 \pi^2 \alpha^{-1} -11\alpha  & \hbox{EM},
\cr}
   \end{equation}
and thus
 \begin{equation}
 g(\alpha)=
 \cases{
  -2  &\hbox{scalar}, \cr  \noalign{\smallskip}
 20 \pi^2 \alpha^{-2} -44 &\hbox{EM}.
 \cr}
 \end{equation}
 Therefore, there is a discrepancy that has nothing to do with a 
bad cutoff but is inherent in either the quantum field theory or 
the basic physics of torque.
Note that the anomaly has a constant sign as a function of~$r$, so 
it cannot disappear when the radial integral is evaluated.

In the framework of Eqs.\ (\ref{startclas})--(\ref{endclas}),
the relevant component of pressure satisfies $p=\beta\rho$ with 
$\beta=3$, just as for parallel plates (to which 
(\ref{wedgestress}) 
formally reduces in the limit of small $\alpha$ and large~$r$).
But the associated  energy density does not satisfy 
(\ref{endclas}) (except in that small-$\alpha$ limit);
it is inconsistent with the equation of state.

 \subsubsection{Conclusion} \label{sec:concl}
 Unless some elementary blunder is being made, the violation of the 
torque balance equation (\ref{torquebal}) by the Deutsch--Candelas 
stress tensor indicates some fundamental problem in our understanding 
of vacuum energy.  It is not an artifact of regularization, 
because no cutoff has been introduced in the analysis.
It apparently has nothing to do with divergences or boundary terms 
in the energy, but rather with the true Casimir energy of the bulk 
region.

  If the wedge is made of thin plates, one might argue that the 
force from  outside the wedge, and the corresponding variation in 
energy, must be taken into account.  These quantities are obtained 
by replacing $\alpha$ by $2\pi-\alpha$  and reversing the sign.
 In the scalar case the anomaly $g$ is independent of~$\alpha$, so the 
total anomaly does vanish in that case.
 In the electromagnetic case, however, it does not cancel (unless 
$\alpha=\pi$).
 Our \emph{Mathematica} calculations indicate  the same conclusion 
for the scalar field with $\xi=\frac14$ and the axial cutoff:  The 
anomaly for $\alpha=\frac{3\pi}4 $ is not the same as that for 
$\alpha=\frac\pi4$.

 It has been suggested that the persistent divergence at $r\to0$ 
spoils the argument:  The conclusion is not convincing unless all 
quantities are finite, or at least all infinities are cleanly 
cancelled by considering the exterior of the region along with the 
interior.
 Therefore, we are presently investigating the Deutsch--Candelas 
stress tensors in an ``annular sector''  and its exteriors.
 That is, we consider conducting boundaries at 
$r = r_{\mathrm{min}}>0$, $r=r_{\mathrm{max}}<\infty$, 
 $\theta = 0$ with $r_{\mathrm{min}}<r<r_{\mathrm{max}}\,$, and  
 $\theta = \alpha$ with $r_{\mathrm{min}}<r<r_{\mathrm{max}}\,$, 
 and allow the last boundary to move.  
 This model has been studied in various ways in \cite{SahT} and 
\cite{MWK}, but those works do not answer all the questions we need 
to ask.
 The divergence at $r=0$ is now removed
 (and would be independent of $\alpha$ anyway). However, new 
divergences are now introduced by the curved boundaries.
 The usual leading-order surface divergences will cancel 
 (in the force) between the 
inside and outside of the wedge surfaces (and between electric and 
magnetic terms in the EM case); we expect them to be nonanomalous 
anyway, on the basis of \cite{EFM}.
 There are also higher-order divergences associated with the 
curvature of the boundary.  The leading such term will cancel, 
exterior of the annulus against the interior.  
 Also, corner energies are independent of~$\alpha$.
 The crux of the 
problem is how the residual bulk Casimir term depends on~$\alpha$.
 One remaining complication is that in the presence of wedge 
boundaries, stress tensors in general are not diagonal; the 
possibility of a nonzero shear force on the curved sides must be 
considered.

 \bigskip
 \begin{acknowledgments}
 We thank Kim Milton, Martin  Schaden, and Lev Kap\-lan for 
comments.
 This research is supported by National Science Foundation Grant
  PHY-0968269. 
 F.D.M. thanks the Mathematics Department of Texas A\&M University 
for renewed hospitality while the manuscript was being written.
\end{acknowledgments}
 \goodbreak

 \end{document}